\documentclass{article}
\usepackage[T1]{fontenc} 
\usepackage[utf8]{inputenc} 
\usepackage{amsmath,cite,url}
\usepackage{graphicx}
\usepackage[lbd]{ismir}

\usepackage{color}

\usepackage[firstpage]{draftwatermark}
\definecolor{lightgray}{rgb}{0.9,0.9,0.9}
\definecolor{darkgray}{rgb}{0.4,0.4,0.4}
\SetWatermarkFontSize{12pt}
\SetWatermarkScale{1.1}
\SetWatermarkAngle{90}
\SetWatermarkHorCenter{202mm}
\SetWatermarkVerCenter{170mm}
\SetWatermarkColor{darkgray}
\SetWatermarkText{Late-Breaking / Demo Session Extended Abstract, ISMIR 2025 Conference}

\title{Latent Granular Resynthesis using Neural Audio Codecs}

\multauthor
{Nao Tokui \hspace{1cm} Tom Baker}
 {Neutone \hspace{1cm} Qosmo\\
{\tt\small \{tokui,tom\}@neutone.ai}}

\def\authorname{N.Tokui and T.Baker}

\usepackage[bookmarks=false,pdfauthor={\authorname},pdfsubject={\pdfsubject},hidelinks]{hyperref}

\begin{document}

\maketitle
%
\begin{abstract}
We introduce a novel technique for creative audio resynthesis that operates by reworking the concept of granular synthesis at the latent vector level. Our approach creates a "granular codebook" by encoding a source audio corpus into latent vector segments, then matches each latent grain of a target audio signal to its closest counterpart in the codebook. The resulting hybrid sequence is decoded to produce audio that preserves the target's temporal structure while adopting the source's timbral characteristics. This technique requires no model training, works with diverse audio materials, and naturally avoids the discontinuities typical of traditional concatenative synthesis through the codec's implicit interpolation during decoding. We include supplementary material\footnote{\href{https://github.com/naotokui/latentgranular/}{\tt github.com/naotokui/latentgranular/}}, as well as a proof-of-concept implementation to allow users to experiment with their own sounds.
\footnote{\href{https://huggingface.co/spaces/naotokui/latentgranular/}{\tt huggingface.co/spaces/naotokui/latentgranular/}}
\end{abstract}

\begin{figure}[h]
    \centering
    \includegraphics[width=0.9\linewidth]{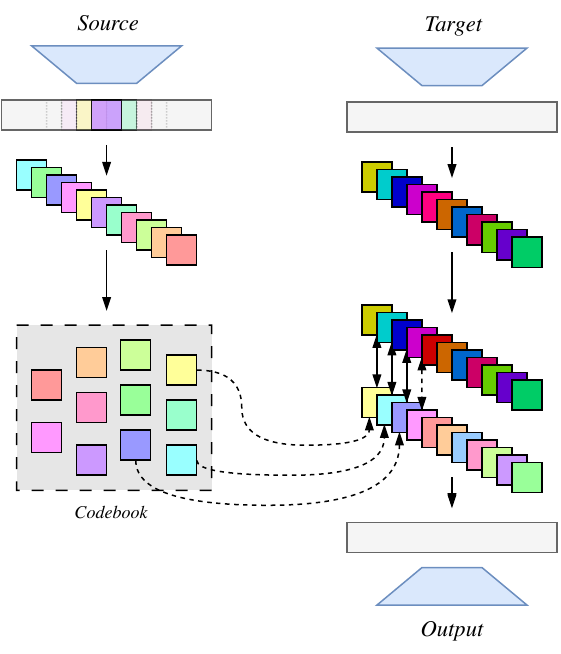} 
    \caption{A process overview. A source audio corpus is encoded to create a codebook of latent vectors. A target audio is encoded, and for each of its vectors, the closest match from the codebook is found to create a new latent sequence, which is then decoded to create the output audio.}
    \label{fig:diagram}
\end{figure}
\section{Background}
Classical granular synthesis\cite{granular}, pioneered in the 1970s, operates by decomposing audio into small fragments or "grains" (typically 1-100ms) which can then be manipulated and recombined to create new textures. Concatenative synthesis (musical mosaicing)\cite{mosaicing} extends this concept to longer segments, focusing on intelligent selection and concatenation of audio units based on acoustic similarity. While these classical techniques have proven valuable for timbre matching and creative sound design, they often suffer from audible discontinuities at grain boundaries due to the discrete nature of audio concatenation.

Recent machine learning approaches have developed upon the foundations of these classical techniques. Autoencoder (AE) based timbre transfer techniques\cite{UMTN, RAVE} offer compelling approaches to generating hybrid sounds by transferring timbral characteristics from one audio source onto another's structure. Similarly, Bitton et al.'s Neural Granular Synthesis\cite{neuralgranular} trains a Variational Autoencoder (VAE) on sound corpora and instead uses granular latent space sampling to generate novel outputs. However, these training-based methods require substantial time and datasets for each corpus, limiting accessibility, and immediate experimentation. Alternatively, "The Concatenator"\cite{concatenator} optimises the concatenative synthesis approach using Bayesian inference with particle filtering for real-time corpus window selection, but still maintains the characteristic granular sound of waveform-domain concatenation.

Neural audio codecs have emerged as powerful tools for high-fidelity audio compression and generation, often employing AE architectures with Residual Vector Quantization (RVQ) to encode audio into compact latent representations while preserving perceptual quality\cite{soundstream, encodec, descript}. Originally designed for efficient audio encoding, adaptations of these codec models have found themselves utilised within latent diffusion models, such as Stable Audio's proprietary VAE\cite{stableaudio, stableaudio2, stableaudioopen} and Diff-a-Riff's Consistency Autoencoder (CAE) Music2Latent\cite{diffariff, dar2, music2latent, music2latent2}. The resulting latent vectors from these pre-trained models provide a far more tractable and computationally efficient medium for audio generation and manipulation compared to raw waveforms.

\section{Methodology}
Our method, illustrated in figure \ref{fig:diagram}, addresses the key limitations of existing approaches by leveraging the compact representations of pre-trained neural audio codecs to create a training-free latent granular framework. By operating in the latent space of these codecs, we eliminate the need for corpus-specific model training while gaining access to high-quality audio representations that naturally interpolate during decoding. This approach combines the modularity and creative flexibility of classical granular/concatenative synthesis with the seamless audio quality achievable through neural compression, enabling immediate experimentation with any source material. 

The method consists of three main stages: codebook generation from source audio, target matching through latent similarity, and reconstruction via decoding.

\subsection{Codebook Generation}
Our approach begins by creating a granular codebook from a source audio corpus through systematic encoding and segmentation. We encode the entire corpus using a pre-trained codec model, then segment the resulting latent representations into "grains" - collections of neighbouring latent vectors that form the basic units of our codebook.

The segmentation process offers two key parameters for creative control. \emph{Grain size} determines how many consecutive latent vectors form each grain (typically 1-5), with the optimal length depending on the source material and desired effect: percussive sounds benefit from smaller grains to capture transient details, while harmonic sounds work better with larger, more consistent windows. \emph{Stride} controls the overlap between consecutive grains - smaller strides provide greater coverage through overlapping segments, while larger strides force more diversity between grains.

The codebook generation process can accommodate diverse source materials, from single instruments to multi-instrumental compositions and non-musical sounds. Multiple codebooks can be created and combined later for modular sound design approaches, allowing artists to blend characteristics from different sources or create layered timbral palettes. 

To expand codebook coverage, we can augment the source data by applying audio effects such as pitch shifting, time stretching, or gain before encoding, effectively broadening the codebook's representation of pitch and timbral space. This augmentation strategy is particularly valuable when working with more limited source material, as it can generate variants that fill gaps in the timbral space.

\subsection{Target Matching}
For any given target audio signal, we apply the same encoding and segmentation process, ensuring we mirror the grains size of the codebook. Each target grain is then matched against the source codebook using cosine similarity as our distance metric.

The sampling strategy for selecting codebook vectors provides key creative control. We sample based on a softmax over negative cosine distances, controlled by a temperature parameter $\tau$:

$$ P(\text{select } B_i) = \frac{\exp(-D_{\cos}(A, B_i) / \tau)}{\sum_{j} \exp(-D_{\cos}(A, B_j) / \tau)} $$

where $D_{\cos}(A, B_i)$ represents the cosine distance between target grain $A$ and codebook grain $B_i$. Lower temperatures provide more faithful timbral matches while closely preserving target structure, whereas higher temperatures introduce randomness and diversity.

Similar to codebook generation, audio effects can be applied to the target signal before encoding to influence the selection process. For example, formant shifting the target can alter which grains are selected from the codebook, creating different timbral mappings while maintaining the target's temporal structure.

We are also exploring alternative matching strategies, such as training a lightweight MLP to extract high-level features from the latent such as pitch, loudness, and timbral descriptors. This approach could enable more nuanced control over which aspects of the source material influence the matching process.

\subsection{Reconstruction}
The final step involves concatenating the selected grain sequence and passing it through the neural audio codec's decoder to generate continuous audio output. This final upsampling performed by the codecs decoder implicitly interpolates between the grains, ensuring a consistent quality of audio output.

\subsubsection{Realtime Capabilities}

Note that the entire grain matching process is completely non-autoregressive. Therefore with an appropriate fast, causal neural audio codec, this whole process can be completely streamable, with latency determined by the codec's inference itself and the grain size.
\section{Conclusion}
We have presented a novel technique that leverages neural audio codecs for creative granular resynthesis, enabling high fidelity, versatile, no-training timbre transfer. By operating in the latent space of pre-trained codecs, our approach achieves smooth timbral blending while preserving the structural characteristics of target audio signals.

In essence, we are creatively "abusing" compression technology originally designed for efficient audio encoding, repurposing it for artistic expression and novel sound generation. The method's strength lies in its simplicity and immediate accessibility — requiring only a source corpus and a target signal to generate compelling hybrid sounds. The ability to work with diverse source materials and adjust granularity provides artists and researchers with powerful tools for creative exploration.

\bibliography{ISMIRtemplate}

%
%
%
%
%

\end{document}